\documentclass[pra,aps,amsmath,amssymb,superscriptaddress,twocolumn,longbibliography]{revtex4-1}
\usepackage{graphicx,multirow}
\usepackage{tikz}
\usepackage{amsmath}
\usepackage{color,outlines}
\usepackage[english]{babel}
\usepackage{amsthm}
\usepackage[sc,osf]{mathpazo}
\usepackage{hyperref}
\usepackage{physics}
\usepackage{outlines}
\usepackage{float}

\definecolor{green2}{RGB}{0,100,0}
 
\newcommand{\e}{\mathbb{E}}

\begin{document}

\title{Multipartite Entanglement Measure : Genuine to Absolutely Maximally Entangled}

\author{Rahul V}
\email{rahulsomasundar@gmail.com}
\author{S. Aravinda}
\email{aravinda@iittp.ac.in}
\affiliation{Indian Institute of Technology Tirupati, Tirupati, India~517619}

\begin{abstract}
Multipartite entanglement is a fundamental aspect of quantum mechanics, crucial to advancements in quantum information processing and quantum computation. Within this field, Genuinely Multipartite Entanglement (GME), being entangled in all bipartitions, and Absolutely Maximally Entanglement (AME), maximally entangled in all bipartitions,  represent two significant types of entanglement with diverse applications. In this work, we introduce a new measure called the GME-AME multipartite entanglement measure, with a non-zero value representing the GME states and the maximum value is reached only by the AME states. The measure is applied to study the multipartite entanglement of four partite systems using the operator to state mapping, and the four partite permutation qutrit states are classified according to the measure. With various examples, we show that our measure is robust in classifying the four partite entangled states. 

\end{abstract}

\maketitle

\section{Introduction} 

Quantum entanglement is a quintessential property of a joint quantum system necessary for various information-theoretic advantages and quantum computation. Entanglement between two parties, referred to as bipartite entanglement, is theoretically well-studied and operationally quantified using various information-theoretic tasks. In these tasks, it can be cleraly shown the advantage of bipartite quantum entanglement~\cite{einstein1935can,horodecki2009quantum} over the any classical resources~\cite{bennett1993teleporting,bennett1984quantum}. Many questions in the bipartite mixed state entanglement persist; the bipartite pure state entanglement is well established. In many-body quantum theory, even though the system consists of many quantum particles, most of the existing study divides the entire many-body system into a bipartition and studies it as a high dimensional bipartite system~\cite{CalabreseCardy2004}. The importance of entanglement and its structure enables important studies in quantum many-body systems, which in turn shape the fields of condensed matter theory and quantum gravity~\cite{van2017lectures,amico2008entanglement}.

The multipartite entanglement theory involves the entanglement shared among three or more subsystems. Unlike bipartite systems, multipartite systems exhibit a much richer and more intricate structure of entanglement, characterized by different forms and classifications. For any quantum computational advantage over classical computation, multipartite entanglement is necessary, but it's not sufficient. For example, in measurement-based quantum computation~\cite{raussendorf2001one}, only certain classes of multiparty entangled states are useful as resource states. For any clear understanding of the advantages of quantum computation and for the construction of successful algorithms, a deeper understanding of multiparty entanglement is essential. 

The bipartite entanglement theory cannot be linearly lifted to multiparty level. There are various issues like defining the maximal entangled state~\cite{guhne2009entanglement}, classification of multipartite entangled states under LOCC~\cite{dur2000three,Chitambar2014}, limitations of entanglement measures for multipartite systems~\cite{eltschka2012multipartite}, complexities in entanglement distillation and manipulation~\cite{PhysRevA.62.062314}, and the challenges associated with extending Bell inequalities and nonlocality arguments to multipartite systems~\cite{Seevinck2017,Brunner2014}.

Genuine multipartite entanglement (GME) and Absolutely Maximally entangled (AME) are the two main types of multipartite entanglement that we are interested in this work. Genuine multipartite entangled state that cannot be factored across any bipartition~\cite{collins2002bell,uffink2002quadratic,seevinck2008partial}, whereas absolutely maximally entangled state exhibits maximum entanglement in all the possible bipartitions~\cite{helwig2012absolute}. 
 The construction of AME states has been explored through various mathematical approaches, such as combinatorial designs, stabilizer codes, and tensor networks~\cite{clarisse2005entangling,goyeneche2018entanglement,goyeneche2015absolutely,rather2023absolutely}.
 The existence of AME states for a general number of systems and local dimensions still remains open, indicating ongoing challenges in characterizing these intriguing quantum states~\cite{huber2020table}. It has been proven that there are no AME states in the four-partite qubit  case~\cite{Higuchi2000}.

Quantifying GME is crucial for understanding its properties and applications~\cite{ma2011measure}. Various measures have been developed to characterize GME states. One widely-used approach is the geometric measure of entanglement, which has been the focus of extensive research efforts ~\cite{shimony1995degree,barnum2001monotones,Wei2003,xie2021triangle,xie2022triangle,xie2024multipartite,mishra2024geometric}. The analysis of density matrix elements also plays a significant role in examining the separability and classification of different GME states ~\cite{Guhne2010}. Numerous studies have employed concurrence-based~\cite{hill1997entanglement,puliyil2022thermodynamic} methods to charecterise multipartite entanglement structures ~\cite{ma2011measure,xie2021triangle,coffman2000distributed}. Several other entanglement measures have been developed to understand multipartite entanglement  ~\cite{wong2001potential,meyer2002global,walter2013entanglement,carvalho2004decoherence}. The  Concentratable Entanglement generalizes various existing measures~\cite{beckey2021computable} in which
its operational meaning is tied to the probability of finding Bell pairs between qubits across different copies of the state.

GME states find applications in quantum sensing ~\cite{giovannetti2011advances}, quantum cryptography ~\cite{proietti2021experimental}, and in quantum metrology ~\cite{Hyllus2012}. In addition, they play a significant role in quantum networks, where their entangled nature allows for efficient entanglement distribution and resource sharing among multiple nodes ~\cite{Liu2020}. AME states are equivalent to error correcting codes~~\cite{raissi2018optimal} and have application in secrete sharing~\cite{helwig2012absolute}, quantum repeters~\cite{alsina2021absolutely} and also in quantum gravity models~\cite{pastawski2015holographic}. Overall, multipartite states' versatile applications underscore their importance in the ongoing development of quantum technologies.

In this work, we construct a measure that can detect both GME and AME, and we term it the GME-AME multipartite entanglement measure. The non-zero value of the GME-AME measure implies the existence of GME states, and the maximum value is reached only by the AME states. As a specific application, we consider the four partite system and studied the GME-AME measure quantitatively. Using the operator state mapping, we construct a class of four party entangled states and calculate the GME-AME measure for this class. We compare the GME-AME with other independent GME measures and AME measures. By using the mapping between permutation operator to permutation states, the entanglement for four party qutrit is studied.

The paper is organized as follows: In Sec.~(\ref{sec:multi}), the multipartite entanglement is introduced and the GME-AME measure is constructed. A brief introduction to other measures is also provided. The operator to state mapping for a four-party system is developed, and its corresponding entanglement measure is calculated in Sec.~(\ref{sec:oper}. For four qubit systems, the GME-AME measure, along with other measures, are calculated in Sec.~(\ref{sec:fourqu}) and for permutation states in Sec.~(\ref{sec:permu}). Finally concluded in Sec.~(\ref{sec:conclu}). 

 \section{Multipartite entanglement measures \label{sec:multi}}

Let $\mathcal{H}_d$ denote the $d$ dimensional Hilbert space of a single system, and the $n$ partite Hilbert space is denoted as $\mathcal{H}_d^{\otimes n} \equiv \mathcal{H}_d \otimes \mathcal{H}_d \otimes \cdots \otimes \mathcal{H}_d$.  The pure bipartite state $\ket{\psi}_{12}$ is called entangled if it cannot be written in product form as $\ket{\psi}_{12} = \ket{\phi}_1 \otimes \ket{\varphi}_2$.   The complexity of characterizing the nature of entanglement increases as the number of parties increases beyond two. There are distinct ways of defining, quantifying and characterizing multipartite entanglement. In this work, we are interested in two variants of multipartite entangled states, genuinely multipartite entangled (GME) and absolutely maximally entangled (AME) pure states. 

An $n$ partite state is fully separable if it can be written as a pure product state $\ket{\Psi} = \ket{\psi_1}_1 \otimes \cdots \ket{\psi_n}_n$, otherwise it is entangled.  If the state $\ket{\Psi}$ is entangled but remains in product form $\ket{\Psi} = \ket{\phi}_A \otimes \ket{\varphi}_{A^c}$ in any bipartition $A/A^c$, then $\ket{\Psi}$ is called biseparable. The state $\ket{\Psi}$ is called a GME if it is not biseparable, it is entangled in all bipartitions.     The state $\ket{\Psi}$ is AME if the reduced state $\rho_A$ of any bipartition $A/A^c$ is a maximally mixed state i.,e $\rho_A \propto I \quad \forall A$.  All AME are GME but the converse is not true.  The relaxed version of AME is the $k$ -uniform states, which is defined as all the reduced states of $k$ partites are proportional to the idenity. The AME corresponds to $k=\lfloor  n/2 \rfloor$. A non-AME $k$ uniform state could be non-GME.  A simple example to demonstrate the differences is that a three-partite GHZ state is AME and also GME, while the three-partite W state is GME but not AME, and a $n$-partite GHZ state is GME but not AME, but is 1-uniform.

As we have explained in the Introduction, there are many measures and witness for a GME and various constructions for AME states. Here we are interested to construct a measure that will witness the GME and also the AME, and we call it the \textit{GME-AME} measure.   The GME-AME measure $\e(\ket{\Psi})$ should satisfy the following conditions: 
\begin{enumerate}
\item $ \e(\ket{\Psi}) = 0$ for all biseparable states. This implies that if $\e(\ket{\Psi}) > 0$, the state $\ket{\Psi}$ is GME. 
\item $ \e(\ket{\Psi}) = 1$ if and only if the state $\ket{\Psi}$ is AME. 
\end{enumerate}
We propose the following measure that satisfies the above conditions: 
\begin{widetext}
\begin{equation}
\e(\ket{\Psi}) =  \left(\frac{d}{d-1}\right)^{m_1} \left(\frac{d^2}{d^2-1}\right)^{m_2} \cdots \left(\frac{d^{\frac{n}{2}}}{d^{\frac{n}{2}}-1}\right)^{m_k}  \prod_{i_1=1}^{m_1} \left(1- \Tr \rho_{i_1}^2\right) \prod_{i_2=1}^{m_2} \left(1- \Tr \rho_{i_2}^2\right) \cdots  \prod_{i_k=1}^{m_k} \left(1- \Tr \rho_{i_k}^2\right).
\label{AME_GME_equ}
\end{equation} 
\end{widetext}
Here $m_l$ is the total number of all possible bipartitions of the reduced state of $l$ partite and $\rho_{i_l}$ is a reduced density matrix of $l$ partite with $ l \in \{1,2,\cdots \lfloor\frac{n}{2} \rfloor \}$. 
Here $m_l$ is the total number of all possible bipartitions of $l$ partite reduced state and $\rho_{i_l}$ is an $l$-partite reduced density matrix with $ l \in \{1,2,\cdots \frac{n}{2}\}$.

  The reduced state $\rho_{i_l}$ will be the pure state if the corresponding bipartition is separable, resulting in a biseparable state $\ket{\Psi}$ hence $\e(\ket{\Psi}) = 0$, and implies that for any  $ \e(\ket{\Psi} >0 $ state is GME.  $\e(\ket{\Psi})=1$  if and only if   $\ket{\Psi}$ is an AME follows from the fact that if the state $\ket{\Psi}$ is AME, the reduced density matrix of all the bipartitions will be maximally mixed and hence the measure $\e(\ket{\Psi})$ reaches the maximum values $1$ (the constants are normalized accordingly), conversely the purity $1-\tr (\rho_i^2)$ is maximum only for the corresponding maximally mixed state $\rho_i \propto I$.

 The two unitary operators $U$ and $U^\prime$ are said to be LU invariant if  $U^\prime = (u_1 \otimes u_2 \otimes \cdots \otimes u_n) U ( u_1^\prime \otimes u_2^\prime \otimes \cdots \otimes u_n^\prime)$. The GME-AME measure $\e(\ket{\Psi})$ is invariant under the local unitary transformation. For any bipartition, the spectrum of the reduced state is LU invariant and hence the measure $\e(\ket{\Psi})$ is also LU invariant, and it is  non-increasing on average under LOCC~\cite{beckey2021computable}. 

In this work we are mainly interested in two other related measures for comparision. The first one is due to Scott~\cite{scott2004multipartite}, which we call here as
the Scott measure~\cite{scott2004multipartite} and  is defined as follows : 
\begin{equation}
    \mathbb{S}ct_k(\ket{\psi}) := \frac{d^k}{d^k-1} \left( 1-\frac{k! (n-k)!}{n!} \sum_{\abs{S} = k} \Tr \rho_s^2 \right)
    \label{Scott_equ}
\end{equation}
where $k = 1,2,\cdots ,\lfloor n/2 \rfloor $, $\lfloor  \rfloor$ denotes the lowest integer part and $S \subset \{1,2,\cdots ,n\}$. 

Some of the properties of $\mathbb{S}ct_k$ 
\begin{enumerate}
    \item $0\leq \mathbb{S}ct_k \leq 1 \quad \forall k$
    \item $\mathbb{S}ct (\ket{\psi}) = 0$ if and only if $\ket{\psi}$ is a pure product state, i.e., $\ket{\psi} = \otimes_{j=1}^n \ket{\psi_j}$. 
    \item $\mathbb{S}ct_k (\ket{\psi}) = 1$ if and only if $\ket{\psi}$ is $k$-uniform, that is all $k$ partite reduced state are maximally mixed. 
\end{enumerate}

\begin{figure}
     \begin{tikzpicture}

\begin{scope}[xscale=2.5, yscale=2.5] 

\coordinate (A) at (0, 0, 0); 
\coordinate (B) at (2, 0, 0); 
\coordinate (C) at (1.8, 2, 2); 
\coordinate (D) at (1.6, 0.2, 1.8); 

\coordinate (P1) at (0.6, 0.33, -0.1); 
\coordinate (P2) at (0.97, 0.03, -0.84); 

\draw[thick] (A) -- (B) -- (C) -- cycle; 
\draw[thick] (A) -- (D); 
\draw[thick] (B) -- (D);
\draw[thick] (C) -- (D);

\node[below left] at (A) {Qubit 4}; 
\node[below right] at (B) {Qubit 2}; 
\node[above] at (C) {Qubit 3}; 
\node[below] at (D) {Qubit 1}; 

\draw[dotted] (A) -- (P1);
\draw[dotted] (C) -- (P1);
\draw[dotted] (D) -- (P1);
\draw[dotted] (D) -- (P2);
\draw[dotted] (C) -- (P2);
\draw[dotted] (B) -- (P2);
\draw[fill=black] (P1) circle (0.02); 
\draw[fill=black] (P2) circle (0.02); 

\draw[thick, black] (1,0.26) circle (0.5); 
\fill[white] (1,0.26) circle (0.5);

\draw[thick] (A) -- (D); 
\draw[thick] (B) -- (D);
\draw[thick] (C) -- (D);

\draw[dotted] (A) -- (P1);
\draw[dotted] (C) -- (P1);
\draw[dotted] (D) -- (P1);
\draw[dotted] (D) -- (P2);
\draw[dotted] (C) -- (P2);
\draw[dotted] (B) -- (P2);
\draw[fill=black] (P1) circle (0.02); 
\draw[fill=black] (P2) circle (0.02); 

\fill[black] (1.7, 0.95, 1.5) ellipse (0.05 and 0.04); 
\fill[black] (1.5, 1.05, 1.8) ellipse (0.05 and 0.04); 
\fill[black] (2.4, 1.5, 2.5) ellipse (0.05 and 0.04);
\fill[black] (2.29, 0.9, 2.4) ellipse (0.05 and 0.04);
\begin{scope}[rotate around={120:(1.7,0.95,1.5)}] 
    \fill[black] (1.7, 0.95, 1.5) ellipse (0.06 and 0.046); 
\end{scope}

\begin{scope}[rotate around={90:(1.5, 1.05, 1.8)}] 
    \fill[black] (1.5, 1.05, 1.8) ellipse (0.06 and 0.046); 
\end{scope}

\begin{scope}[rotate around={120:(2.4, 1.5, 2.5)}] 
    \fill[black] (2.4, 1.5, 2.5) ellipse (0.06 and 0.04); 
\end{scope}

\begin{scope}[rotate around={90:(2.29, 0.9, 2.4)}] 
    \fill[black] (2.29, 0.9, 2.4) ellipse (0.06 and 0.046); 
\end{scope}

\draw[->] (1.7, 0.95, 1.5) -- (-0.25, 0.2, -2); 
\draw[->] (1.5, 1.05, 1.8) -- (-0.25, 0.15, -2);

\draw[->] (2.4, 1.5, 2.5) -- (1, 0.2, -1.8);

\draw[->] (2.29, 0.9, 2.4) -- (2, 0.5, -0.2);

\node[above] at (-0.25, 0.2, -2) {$\gamma_{13}$};
\node[above] at (1, 0.2, -1.8) {$\gamma_{23}$};
\node[above] at (2, 0.5, -0.2) {$\gamma_{12}$};
\end{scope}

\end{tikzpicture}
     
     \caption{Tetrahedron with qubits on its 4 vertices, each face is divided into 3 triangle and its area is denoted as $\gamma_{12},\gamma_{13},\gamma_{23}$ and $\gamma_{ij}$ is the area of triangle by the qubit $i$, qubit $j$ and the mid point of that face} 
     \label{fig:1}
 \end{figure}
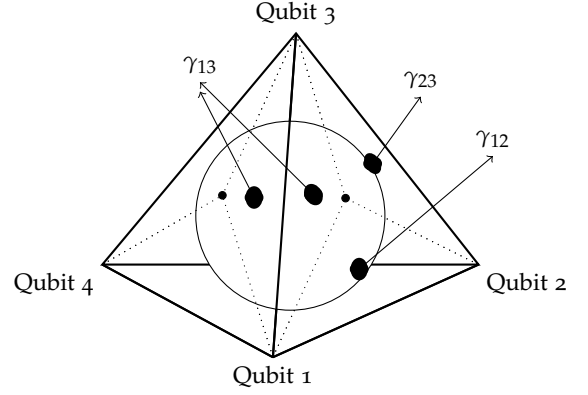

Similarly, the GME measure that we are interested in in this work is based on the triangle measure by Xie et al.~\cite{xie2022triangle}, and extended to the four qubits system in Ref.~\cite{xie2024multipartite}. This measure uses geometric intuition to quantify the genuine entanglement. 
 The  GME measure for four-qubit system is defined using the volume $V$ of a simplex polygon, which is given as : 
\begin{equation}
    V = \frac{\sqrt{2}}{3} S^{1/2} (R_0 R_1 R_2 R_3)^{1/4},
\end{equation}
where the coefficients $R_0, R_1, R_2, R_3$ and $S$ are functions of the pairwise correlations $\gamma_{ij}$, and  $\gamma_{ij}$ are the area of the polygon shown in Fig.~(\ref{fig:1}) are :
\begin{equation}
    \begin{split}
       & R_0 = +\sqrt{\gamma_{12}\gamma_{34}} + \sqrt{\gamma_{13}\gamma_{24}} +\sqrt{\gamma_{14}\gamma_{23}} \\
        & R_1 = -\sqrt{\gamma_{12}\gamma_{34}} + \sqrt{\gamma_{13}\gamma_{24}} +\sqrt{\gamma_{14}\gamma_{23}} \\
         & R_2 = +\sqrt{\gamma_{12}\gamma_{34}} - \sqrt{\gamma_{13}\gamma_{24}} +\sqrt{\gamma_{14}\gamma_{23}} \\
          & R_3 = +\sqrt{\gamma_{12}\gamma_{34}} + \sqrt{\gamma_{13}\gamma_{24}} -\sqrt{\gamma_{14}\gamma_{23}} \\
          & S=2(\gamma_{12}+\gamma_{13}+\gamma_{14}+\gamma_{23}+\gamma_{24}+\gamma_{34})
    \end{split}
\end{equation}
The $\gamma's$ can be obtained by simultaneously solving the following Eq.~(\ref{polygon}).

\begin{equation}
\begin{split}
    \gamma_{ij} + \gamma_{ik} + \gamma_{il} = & E_{i|jkl} \\
     -\sqrt{\gamma_{ij}\gamma_{kl}} + \sqrt{\gamma_{ik}\gamma_{jl}} +\sqrt{\gamma_{il}\gamma_{jk}}  = & \lambda E_{ij|kl},
\end{split} 
\label{polygon}
\end{equation}
where  $E_{ij|kl}$ and  $E_{i|jkl}$ are bipartite and single partite von Neumann entropies. 
The polygon measure $\mathbb{P}$ is defined as
\begin{equation}
    \mathbb{P} = \frac{3^{7/6}}{2} V^{2/3}.
    \label{GME_equ}
\end{equation}
Here $0 < \mathbb{P} \leq 1$ for all GME states and $\mathbb{P} = 0$ for any non-GME states.

\section{Entanglement of four partite system \label{sec:oper}}

Many interesting properties of the four partite systems can be studied by using operator to state mapping, the Choi-Jamialkowski isomorphism, as the operator acting on two systems can be mapped to four partite systems. 
A $d\times d$ matrix (an operator) $A = \sum_{ij} \mel{i}{A}{j} \ketbra{i}{j} \in \mathcal{H}_d$ can be mapped to a state $\ket{A}_{12} = \frac{1}{\sqrt{d}} \sum_{ij} \mel{i}{A}{j} \ket{ij} \in \mathcal{H}_d \otimes \mathcal{H}_d $. 
Similarly, an operator $A = \sum_{i\alpha,j\beta} \mel{i\alpha}{A}{j\beta} \ketbra{i\alpha}{j\beta} \in \mathcal{H}_{d} \otimes \mathcal{H}_{d}$ is mapped to a four partite state $\ket{A}_{1234} = \frac{1}{d}\sum_{i\alpha,j\beta} \mel{i\alpha}{A}{j\beta} \ket{i\alpha j\beta} $~\cite{Zyczkowski2004}. By using the vectorization identity $\ket{XYZ} = (Z^T \otimes X ) \ket{Y}$, the local unitarily (LU) invariant operators $A^\prime = u_1 \otimes u_2 A u_3 \otimes u_4$ is mapped to LU invariant four party state as $\ket{A^\prime} = u_3^T \otimes u_4^T \otimes u_1 \otimes u_2 \ket{A}$, thereby preserving all the entanglement structures. Here $T$ represents full transpose, $\mel{i\alpha}{A}{j\beta} = \mel{j\beta}{A^{T}}{i\alpha}$.

The two-particle reduced density matrix takes a simple form in this representation, easing many calculations. For that, we need to define various reshaping maps of the matrix~\cite{Zyczkowski2004,Bengtsson2007}. The realignment and partial transpose are such reshaping operations that are widely used in the entanglement theory~\cite{chen2002matrix,rudolph2003some} and open quantum systems ~\cite{oxenrider1985matrix,sudarshan1961stochastic}. Recently, these operations are involved in defining the solvable models of many-body systems and have provided a host of analytically solvable properties~~\cite{Akila2016,Bertini2019,SAA2020,ASA_2021,RAA22}. We use the following reshaping of matrices in our work: 
\begin{equation}
    \begin{split}
        \text{Realignment 1:} &\quad \mel{i\alpha}{A}{j\beta} = \mel{\beta\alpha}{A^{R_1}}{ji} \\
        \text{Realignment 2:} &\quad \mel{i\alpha}{A}{j\beta} = \mel{ij}{A^{R_2}}{\alpha\beta} \\
        \text{Partial Transpose 1:} &\quad \mel{i\alpha}{A}{j\beta} = \mel{j\alpha}{A^{T_1}}{i\beta} \\
        \text{Partial Transpose 2:} &\quad \mel{i\alpha}{A}{j\beta} = \mel{i\beta}{A^{T_2}}{j\alpha}. 
    \end{split}
\end{equation}

In this terminology, for a four partite state 
\begin{equation}
    \ket{A}_{1234} = \frac{1}{d}\sum_{i\alpha,j\beta} \mel{i\alpha}{A}{j\beta} \ket{i\alpha j\beta},
\end{equation}
the two particle-reduced density matrix is given as 
\begin{equation}
    \label{alignment}
    \rho_{12} = \frac{AA^\dagger}{d^2},   \quad \rho_{13} = \frac{A^{R_2}A^{R_2\dagger}}{d^2},   \quad \rho_{14} = \frac{A^{T_2}A^{T_2\dagger}}{d^2}.  
\end{equation}
The advantage of this representation is visible by considering the measure $\e{\ket{\Psi}}$ for four parties, 
\begin{equation}
\begin{split}
    \e({\ket{\Psi}}) = &\left(\frac{d}{d-1}\right)^4 \left(\frac{d^2}{d^2-1}\right)^3 \\
   & \prod_{j=1}^4 (1-\tr \rho_j^2) \prod_{r\in 12,13,14} (1-\tr \rho^2_r).
\end{split}
    \label{eq:frm}
\end{equation}
Suppose the two-particle operator $A$ is unitary, $\rho_{12} = \frac{I}{d^2}$, and hence all single particle reduced state $\rho_j = \frac{I}{d} \quad \forall j \in \{1, \cdot ,4\}$, and then the measure $\e(\ket{\Psi})$  reduces to 
\begin{equation}
    \e({\ket{\Psi}}) = \left(\frac{d^2}{d^2-1} \right)^2 (1-\tr \rho_{13}^2 ) (1-\tr \rho_{14}^2).
\end{equation}
For the identity matrix $I = \sum{ij} \ketbra{ij}{ij}$, the realignment is $I^{R_2} = \sum_{ij} \ketbra{ii}{jj}$, hence the state $\rho_{13} = \ketbra{\phi^+}{\phi^+}$, where $\ket{\phi^+} = \frac{1}{\sqrt{d}} \sum_{i} \ket{ii}$, a two particle maximally entangled state~\cite{ASA_2021}. Hence, the state $\ket{I}$ and any LU equivalent to $\ket{I}$ is non-GME for that $\e(\ket{I}) = 0$. Similarly, consider \texttt{swap} $S = \sum_{ij} \ketbra{ji}{ij}$, for which $S^{T_2} = \sum_{ij} \ketbra{jj}{ii}$, hence the state $\rho_{14} = \ketbra{\phi^+}{\phi^+}$, that the state $\ket{S}$ and its LU equivalent  is non-GME, and $\e(\ket{I}) = 0$. This provides an elegant way to construct any four partite GME states. Any diagonal unitary operator $U = \sum_{ij} e^{\theta_{ij}} \ketbra{ij}{ij}$ remains unitary after partial transposition $T_2$, and the state $\rho_{14} = \frac{I}{d^2}$~\ref{alignment}, and hence except the identity operator $I$ and its LU equivalent, $\e(\ket{\Psi}) > 0$. Similarly, except for SWAP and its LU equivalent, any unitary operator after realignment $R_2$ remains unitary, and its corresponding $\ket{\Psi}$ is GME. Such unitary operators are called dual-unitary operators~~\cite{Akila2016,Bertini2019} and have been studied widely in many-body exactly solvable models. Various construction methods of such dual-unitary operators provide a host of four partite GME states.     

\section{Four qubit states \label{sec:fourqu}}
\begin{figure}[h!]
    \centering
    \begin{tikzpicture}
        \begin{scope}[xscale=0.4, yscale=0.4] 
        
            \coordinate (A) at (-2,2); 
            \coordinate (B) at (0,0); 
            \coordinate (C) at (4,2); 
            \coordinate (D) at (1,8); 

            \draw (A) -- (B) -- (C) -- (D) -- (A);
            \draw (B) -- (D);
            \draw (A) -- (C);

            \node[ left] at (A) {\parbox{1cm}{LOCAL \\ $(0,0,0)$}};
            \node[below ] at (B) {\parbox{1cm}{CNOT \\ $\left(\frac{\pi}{4}, 0, 0\right)$}};
            \node[ right, xshift=0.5mm, yshift=0.1mm] at (C) {\parbox{0.1cm}{DCNOT \\ $\left(\frac{\pi}{2}, \frac{\pi}{4}, 0\right)$}};

            \node[above] at (D) {\parbox{2cm}{SWAP \\ $\left(\frac{\pi}{4}, \frac{\pi}{4}, \frac{\pi}{4}\right)$}};
        
        \end{scope}
    \end{tikzpicture}
    \caption{Weyl Chamber representation of LU equivalent parameterized  nonlocal unitary operators, vertices are represented as LOCAL, CNOT, DCNOT, and SWAP gates with their corresponding parameter values.}
    \label{fig:weyl}
\end{figure}
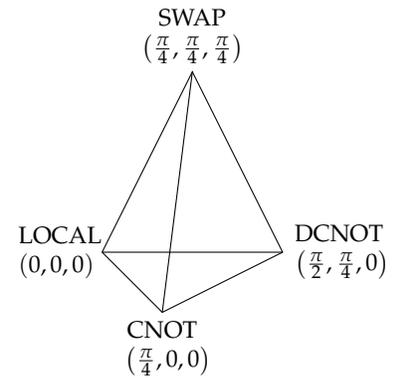

The unitary operator $U$ acting on two qubits is LU equivalent to a three-parameter family of unitary operators $X$ expressed in its canonical called Cartan decomposition as~~\cite{KBG01,KC01,Zhang2003} 
\begin{equation}
\begin{split}
     X &= \exp[-i ( x \sigma_x \otimes \sigma_x + y \sigma_y \otimes \sigma_y + z \sigma_z \otimes \sigma_z)] \\  &= \begin{bmatrix}
        e^{-iz} c_{-} & 0 & 0& -ie^{-iz} s_{-} \\
        0 & e^{iz} c_{+} & -ie^{iz} s_{+} & 0 \\
        0 &-ie^{iz} s_{+} &e^{iz} c_{+} & 0 \\
        -ie^{-iz} s_{-} & 0 &0 & e^{-iz} c_{-}
    \end{bmatrix} 
\end{split}
\end{equation}
where $c_{\pm} = \cos(x \pm  y), s_{\pm} = \sin(x \pm  y))$ and $0\leq \abs{z} \leq y \leq x \leq \pi/4$. Thus the coefficients form a tetrahedron called a Weyl chamber represented in Fig.~(~\ref{fig:weyl})

\begin{figure}[H]
    \centering
    \includegraphics[width=7.3cm]{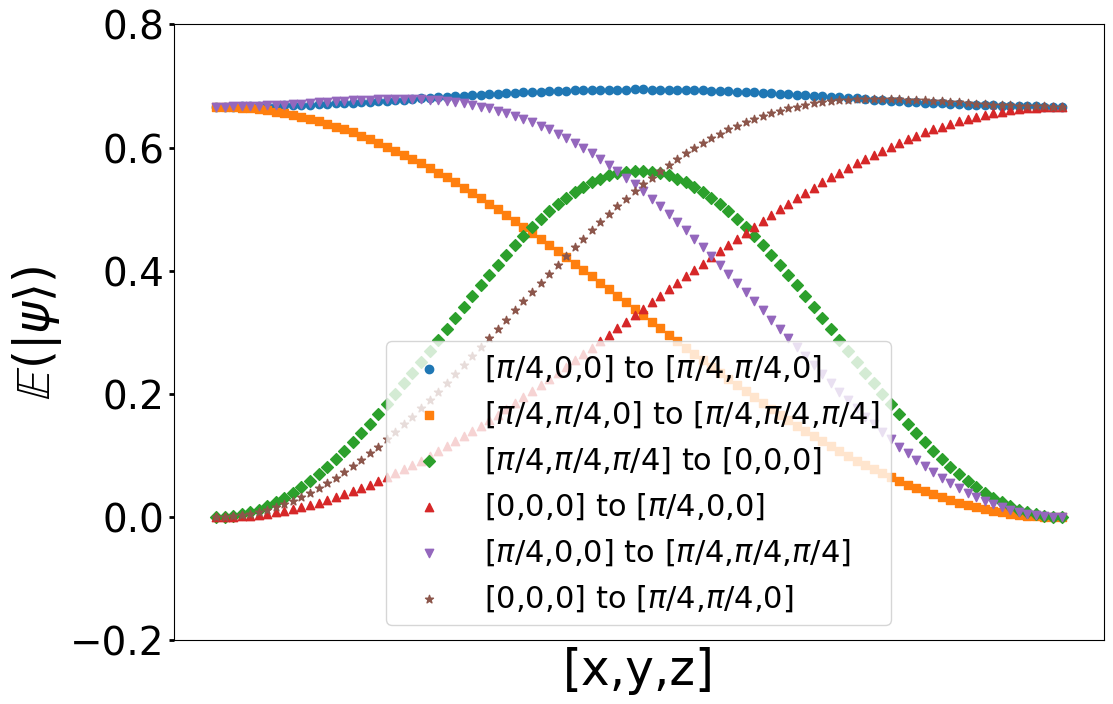}
    \caption{The value of GME-AME measure  $\mathbb{E}(\ket{X})$ varying the Weyl parameters  $[x, y, z]$. It is generated from 100 data points obtained by by varying the values of $[x, y, z]$ satisfying Weyl condition.} 
    \label{fig:quGMEAME}
\end{figure}

The operator state mapping of the unitary operator $X$ to $\ket{X}$ provides a parameterized set of four party qubit states having various entanglement structure. The GME-AME measure $\e(\ket{X})$ can be calculated explicitly for the family of $\ket{X}$ states and given as : 
\begin{widetext}
\begin{equation}
\begin{split}
    \mathbb{E}(\ket{X}) = & \frac{16}{9} \Bigg( 1 - \frac{1}{8} \left[ \left( \cos^2(x - y) + \cos^2(x + y) \right)^2 + \left( \sin^2(x + y) + \sin^2(x - y) \right)^2 
    + \left( 2 \cos(x - y) \cos(x + y) \cos(2z) \right)^2  \right. \\
    & \left. + \left( 2 \sin(x + y) \sin(x - y) \cos(2z) \right)^2 \right] \Bigg) \\
    &  \Bigg( 1 - \frac{1}{8} \left[ \left( \cos^2(x - y) + \sin^2(x + y) \right)^2 + \left( \cos^2(x + y) + \sin^2(x - y) \right)^2 \right. \\
    & \left. + \left( 2 \cos(x - y) \sin(x + y) \sin(2z) \right)^2 + \left( 2 \cos(x + y) \sin(x - y) \sin(2z) \right)^2 \right] \Bigg).
\end{split}
\end{equation}

\end{widetext} 
The GME-AME measure is interesting to consider on the state $\ket{X}$ for the values corresponding to  the edges of the Weyl chamber. For four qubit states, GME-AME can never take the maximum value of 1, as four qubit AME doesn't exist~~\cite{Higuchi2000}. The GME-AME values for the state $\ket{X}$ has been plotted for various values of $[x,y,z]$ in Fig.~(\ref{fig:quGMEAME}). For the single parameter family, the plot ranges for all the values, and for two and three parameter family, we have considered the equal values for all the parameters to satisfy the Weyl condition. 

For the line connecting between LOCAL and SWAP for values $(0,0,0)$ to $(\frac{\pi}{4}, 0,0)$, $X^{T_2}$ is unitary and hence the reduced state $\rho_{14}$ ~\ref{alignment} is maximally mixed, and the corresponding GME-AME value is
\begin{equation}
\mathbb{E}(\ket{X}) = \frac{1 - \cos(4x)}{3},
\end{equation}
for which the maximum value $2/3$ is reached for the value $x=\pi/4$ and the $0$ for the value $x=0$. Similarly, for the edge between SWAP to DCNOT, $(\frac{\pi}{4},\frac{\pi}{4},\frac{\pi}{4})$ to $(\frac{\pi}{4}, \frac{\pi}{4},0)$, the reduced matrix $\rho_{13}$ is maximally mixed and the GME-AME value varies between $0$ and $2/3$ for $z=0$ to $z=\pi/4$, and for other value the expression is given as 
\begin{equation}
\mathbb{E}(\ket{X}) = \frac{\cos(4z) + 1}{3}.     
\end{equation}
Similarly for other values, also we can obtain the simplified expression for the GME-AME measure and are expressed as follows: 
(1) For the vertex CNOT to DCNOT, for the values $(\pi/4,0,0)$ to $(\pi/4,\pi/4,0)$, the GME-AME measure reduces to a single parameter family as : 
\begin{widetext}
\begin{equation}
    \mathbb{E}(\ket{X}) = \frac{2}{10} \left( \cos(4y) - \frac{10}{9} \right) \left( \sin^4 \left( \frac{y + \frac{\pi}{4}}{2} \right) + \cos^4 \left( \frac{y + \frac{\pi}{4}}{2} \right) - 1 \right). 
\end{equation}
\end{widetext}
(2) For the vertex CNOT to SWAP, $(\pi/4,0,0)$ to $(\pi/4,\pi/4,\pi/4)$, the GME-AME measure is 
\begin{widetext}
\begin{equation}
    \mathbb{E}(\ket{X}) = \left( 1.7 \sin^2\left(\frac{\pi}{4} + y\right) \cos^2(2z) \cos^2\left(\frac{\pi}{4} + y\right) - \frac{4}{3} \right)  \times \left( \sin^2(2z) + \sin^2\left(\frac{\pi}{4} + y\right) 
        + \cos^2\left(\frac{\pi}{4} + y\right) - 1 \right).
\end{equation}
\end{widetext}
(3) For the vertex LOCAL to DCNOT, $(0,0,0)$ to $(\pi/4,\pi/4,0)$ 
\begin{widetext}
\begin{equation}
\begin{split}
   &\mathbb{E}(\ket{X}) =  \left( 4.0 \sin^2(x) \sin^2(y) \cos^2(x) \cos^2(y) - 0.75 \right)  \\
   &\left( 0.22 (1 - \cos(2x))^2 (1 - \cos(2y))^2  + 0.22 (\cos(2x) + 1)^2 (\cos(2y) + 1)^2 - 0.88 \cos(2x - 2y) - 0.88 \cos(2x + 2y) - 1.78 \right).
   \end{split}
\end{equation}
\end{widetext}

\begin{figure}[htbp]
    \centering
    \includegraphics[width=.5\textwidth]{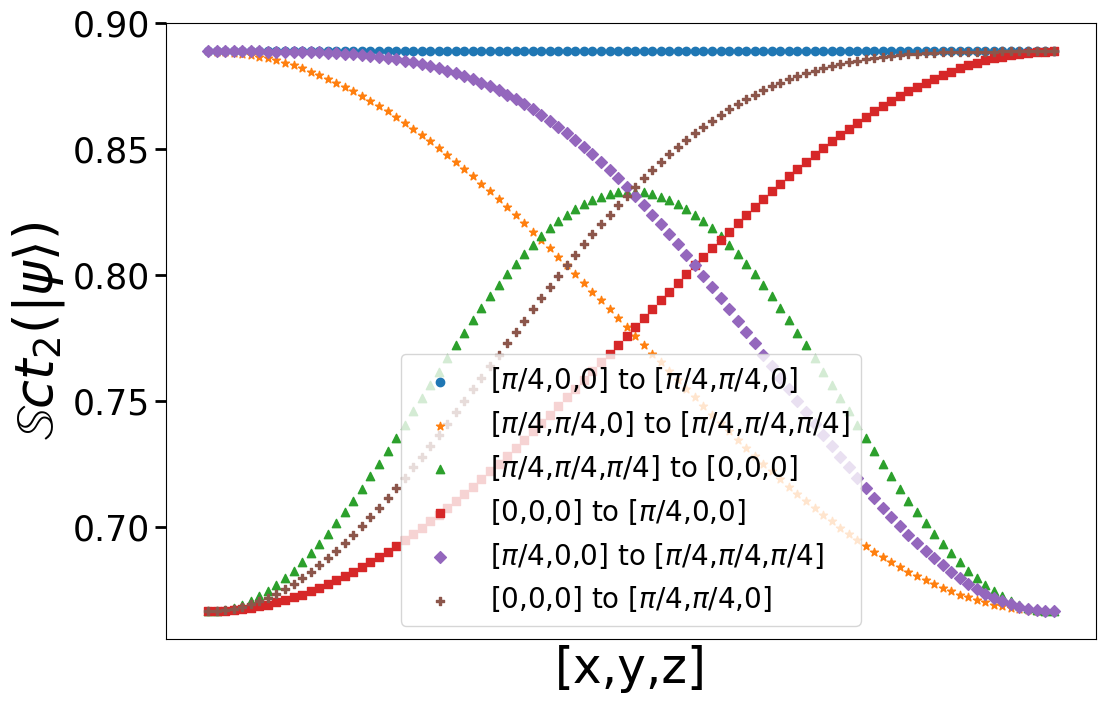} 
    \caption{The Scott measure $\mathbb{S}ct_{2}(\ket{X})$ for qubits, with the plot generated from 100 data points obtained by varying the parameters [x, y, z].} 
    \label{sct_x}
\end{figure}

\begin{figure}[h]
    \centering
    \includegraphics[width=.5\textwidth]{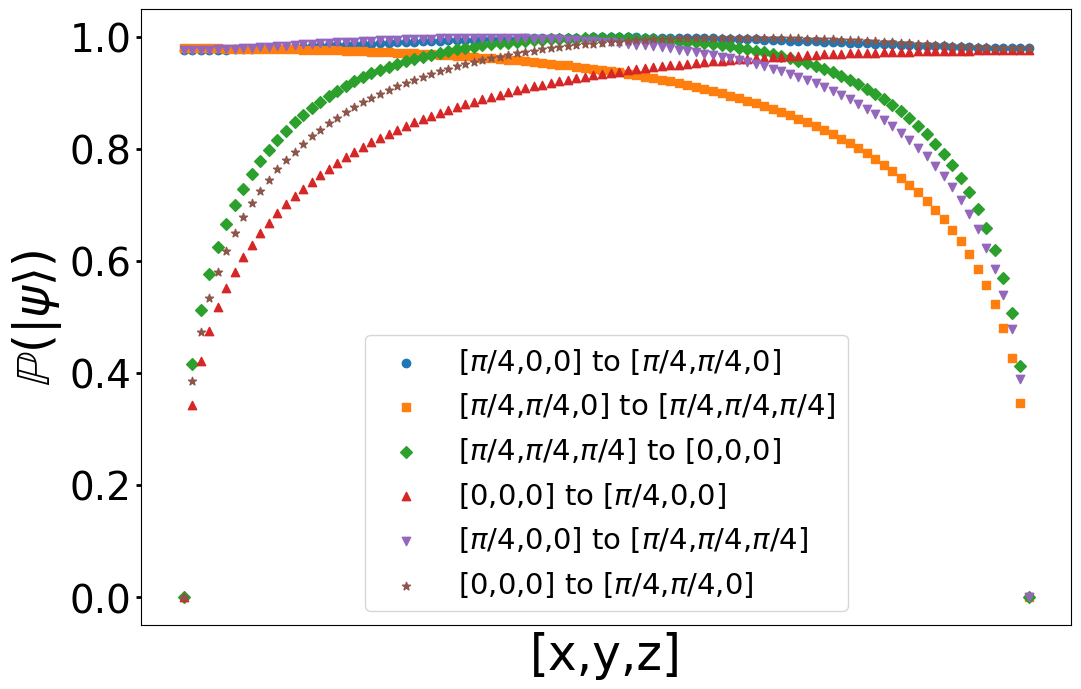} 
    \caption{The Polygon measure  $\mathbb{P}(\ket{\psi})$ of $\ket{X}$.  The state shows $\mathbb{P} = 0$ for the specific parameter sets $ [x, y, z] = [0, 0, 0]$  and [$\pi/4$, $\pi/4$, $\pi/4$], while exhibiting non-zero values for all other parameter combinations.}
    \label{P_X}
\end{figure}

The Scott measure, Eq. (\ref{Scott_equ}) and the Polygon measure in Eq.~(\ref{GME_equ})is calculated for the parameterized state $\ket{X}$. The detailed plots for all the values of $[x,y,z]$ from the Weyl chamber is plotted for Scott measure in Fig.(\ref{sct_x}) and the Polygon measure in the Fig. (\ref{fig:quGMEAME}). The Scott measure can be evaluated analytically, and given as 
\begin{equation}
  \mathbb{S}ct_2(\ket{X})   = \frac{4}{3} - \frac{1}{18} f(x,y,z)
\end{equation}
where 
\begin{equation}
    \begin{split}
        f(x,y,z) &= 2+ 4 \cos^4(x-y) + (-1+ \sin(2x)\sin(2y))^2 \\&+ (1+\sin(2x)\sin(2y))^2 \\&+(\sin^2(x-y) +\sin^2(x+y))^2 + \\&4\cos^2(2z) (\cos^2(x-y)\cos^2(x+y) +\sin^4(x+y)) \\&-(-2 + \cos(4x) + \cos(4y))\sin^2(2z)
    \end{split}
\end{equation}
which shows similar behavior to that of the GME-AME measure but differing the fact that it is unable to detect the non-GME states. The Polygon measure differs predominently with the GME-AME measure. It detects the GME state successfully but fails to detect the AME, as it will reach maximum value even for non-AME states.

For three qubit states, GHZ and W states are two LOCC inequivalent classes~~\cite{dur2000three}. For four qubit states there are continous family of LOCC inequivalent classes divided into nine distinct classes~~\cite{verstraete2002four}. The GME-AME values for the all these states are calculated and is presented in the Appendix.~(\ref{9 different classes}).

\section{Four party permutation states \label{sec:permu}}

\begin{widetext}
  \begin{table}[H]
    \centering
    \begin{tabular}{|c|c|c| |c|c|c| |c|c|c| |c|c|c| |c|c|c| |c|c|c| |c|c|c| |c|c|c|} \hline
        \textbf{No.} & $\e (\ket{P})$ & \textbf{States} & \textbf{No.} & $\e (\ket{P})$ & \textbf{States} & \textbf{No.} & $\e (\ket{P})$ & \textbf{States} & \textbf{No.} & $\e (\ket{P})$ & \textbf{States} & \textbf{No.} & $\e (\ket{P})$ & \textbf{States} & \textbf{No.} & $\e (\ket{P})$ & \textbf{States} \\ \hline
        1  & 0.0000 & 72    &  7  & 0.6667 & 13608 & 13 & 0.7222 & 2592 & 19  & 0.6420 & 5184  &  24 & 0.7176 & 10368 & 29 & 0.7654 & 64800 \\
        2  & 0.4444 & 1296  &  8  & 0.6713 & 10368 & 14 & 0.7346 & 5184  &20 & 0.7894 & 10368 &  25 & 0.8148 & 15552 & 30 & 0.8056 & 5184 \\
        3  & 0.5000 & 864   &  9 & 0.6821 & 10368 & 15 & 0.7407 & 10368 & 21  & 0.6667 & 13608 &  26 & 0.7222 & 2592  & 31 & 0.7901 & 34344 \\
        4  & 0.5093 & 1296  &  10 & 0.6914 & 12960 & 16 & 0.7415 & 25920  & 22 & 0.8133 & 25920 &  27 & 0.8395 & 20736 & 32 & 0.8920 & 2592  \\
        5  & 0.6019 & 1296  &  11 & 0.6944 & 3456  & 17 & 0.7500 & 288   & 23 & 0.8889 & 1296  &  28 & 0.8403 & 1728  & 33 & 1.0000 & 72    \\
        6  & 0.6296 & 5184  &  12 & 0.7160 & 23328 & 18 & 0.7608 & 36288 & & & & & & & & & \\ \hline
        \end{tabular}
    \caption{The GME-AME value $\mathbb{E}(\ket{P}$ for the  permutation qutrit states. It shows 33 classes of  permutation states are there out of $9!=362880$ states }
    \label{E_qutrit_table}
\end{table}
\end{widetext}

In quantum information theory, permutation matrices play a fundamental role in describing the rearrangement of subsystems in composite quantum systems. These matrices are used to model the swapping of qubits or higher-dimensional quantum states (qudits) and are integral in the study of entanglement, error correction, and quantum algorithms.  A permutation matrix is a square unitary matrix that has exactly one entry of 1 in each row and each column, with all other entries being 0.  

Let $P$ represent the permutation matrix of size $d^2 \times d^2$ defined on the bipartite Hilbert space $\mathcal{H}_d \otimes \mathcal{H}_d$, 
\begin{equation}
P\ket{ij} = \ket{m_{ij}n_{ij}}, 
\end{equation}
such that the operator and corresponding permutation state is defined as 
\begin{equation}
\begin{split}
    P & = \sum_{ij} \ketbra{m_{ij}n_{ij}}{ij} \\ 
    \ket{P} & = \sum_{ij} \ket{m_{ij}n_{ij}} \ket{ij}
\end{split}
\label{eq:permu}
\end{equation}
and the corresponding four party state is represented as $\ket{P}$. In this section, we study the entanglement structure of these states for local dimension $d=2$ and $d=3$.  

\begin{figure}[H]
    \centering
    \includegraphics[width=8cm]{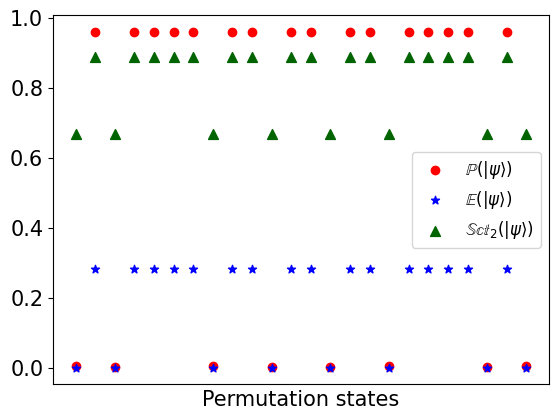}
    \caption{$\mathbb{E}(\ket{P})$, $\mathbb{S}ct_2(\ket{P})$ and $\mathbb{P}(\ket{P})$ of  qubit permutation state.}
    \label{comparison}
\end{figure}

For any dimension $d$ there are $d^2 !$ number of permutation operators can be defined. For $d=2$, there are $24$ permutation operators and corresponding $24$ perumtation states $\ket{P}$. We have calculated the GME-AME measure, Scott measure and Polygon measure and plotted in the Fig. (\ref{comparison}). It can be seen that there are two classes states having minimum and maximum, which is evident from all the three measures. There are $8$ permutation states that are non-GME and remaining $16$ states are GME states maximizing the respective measures.  For GME-AME measure, and for the Polygon measure, all $8$ permutation operators reaches its minimum $0$, and the remaining $16$ permutation states reach the  maximum value $0.3$ for the GME-AME measure 0.7 for Scott measure  and the value $0.96$ for the Polygon measure. 

\begin{figure}[h]
    \centering
    \includegraphics[width=8cm]{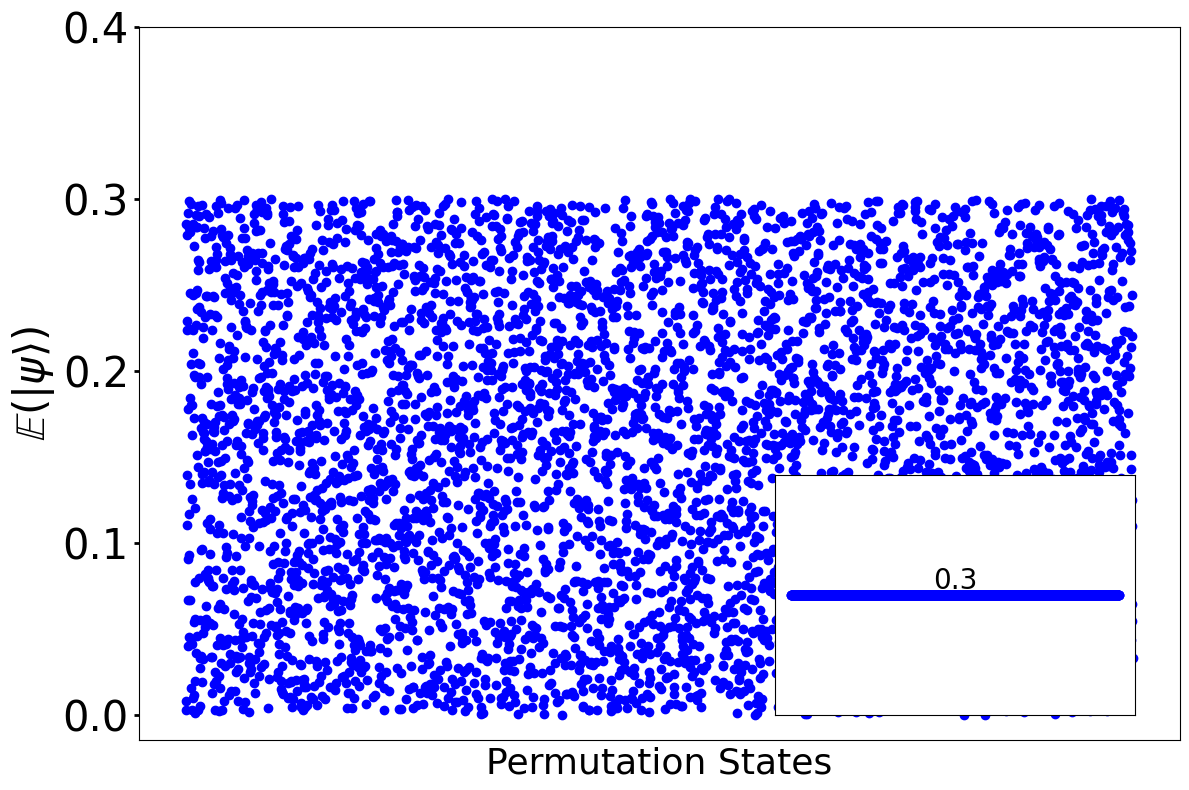}
    \caption{The GME-AME measure $\mathbb{E}(\ket{\psi})$ of enphased permutation state by enphasing the permutation having $\mathbb{E}\ket{\psi}=0$ and the inset one is the enphased with the permutation having $\mathbb{E}\ket{\psi}=0.3$.}
    \label{E_enphase}
\end{figure}

Interesting way to change the entanglement structure is {\it enphasing} the permutation states by introducing the arbitrary phase~~\cite{SAA2020}, 
\begin{equation}
\begin{split}
    P &= \sum_{ij} e ^{i \theta_{ij}} \ket{ij}\bra{l_{ij} k_{ij}} \\
    \ket{P} &= \sum_{ij} e ^{i \theta_{ij}} \ket{ij}\bra{l_{ij} k_{ij}}. 
\end{split}
\end{equation}
For $d=2$, such enphasing for $\theta_{ij} \in \{0,\pi\}$ is considered for a both the class of permutation operators and the variation of various measures is plotted.  For the the permutation state having either $0$ value (non-GME) or minimum value (in case of Scott measure), enphasing produces $\mathbb{E}(\ket{\psi})$ values that range from $0.0$ to $0.3$ Fig :~(\ref{E_enphase} case of GME-AME measure, 0.2 to 0.96 value in case of polygon measure Fig : ~\ref{polygon enphasing}, 0.66 to 0.88 in case of Scott measure Fig : ~\ref{scott enphasing}. In contrast, for the GME states , enphasing yields a constant for all three measures .

\begin{figure}[h]
    \centering
    \includegraphics[width=8cm]{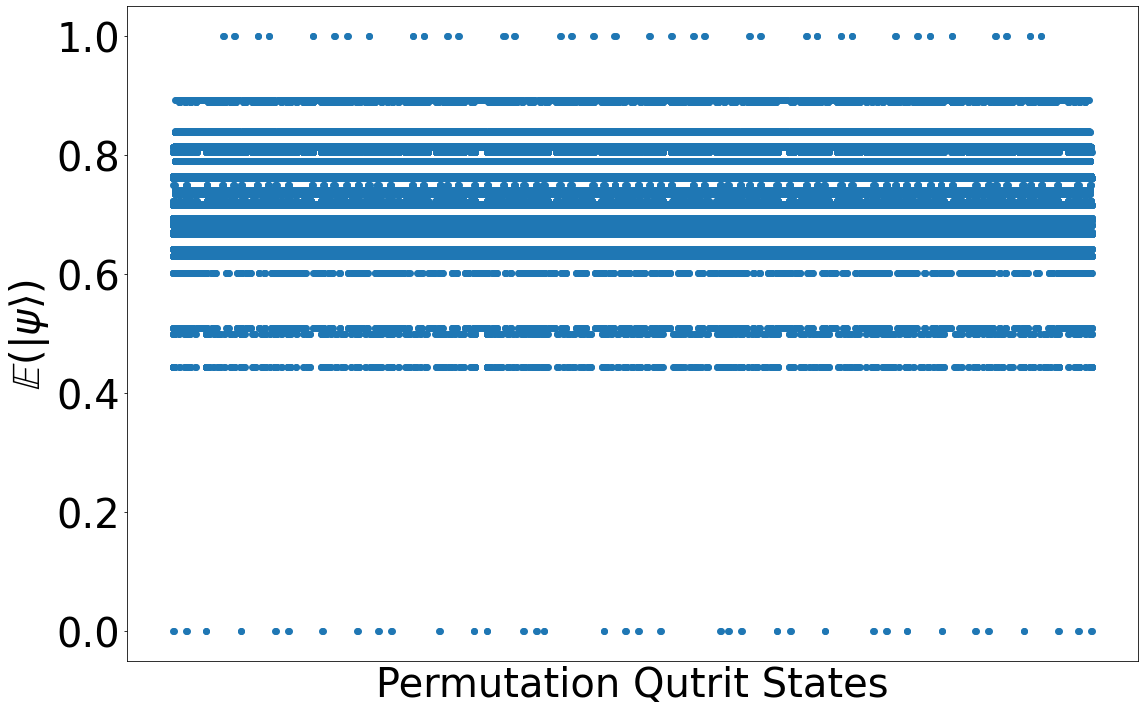}
    \caption{The GME-AME $\mathbb{E}(\ket{P})$ for the permutation of a qutrit state. Among the 9! = 362,880 possible states, 33 distinct $\mathbb{E}(\ket{P})$ values are observed.}
    \label{E_qutrit}
\end{figure}

\begin{figure}[h]
    \centering
    \includegraphics[width=8cm]{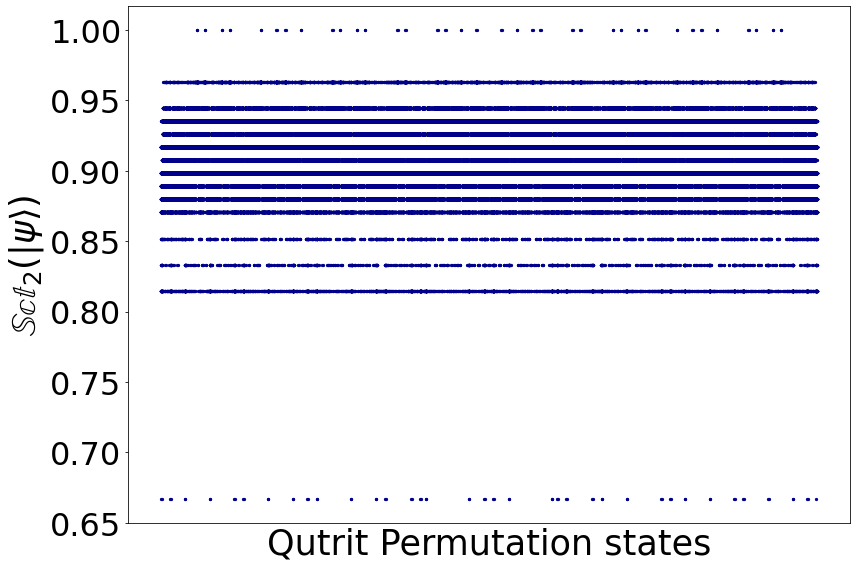}
    
    \caption{The Scott measure $\mathbf{Sct}_{2} (\ket{P})$ for the permutation of a qutrit state. Among the 9! = 362,880 possible states, 16 distinct $\mathbb{E}(\ket{P})$ values are observed.}
    \label{scott_qutrit}
\end{figure}

For $d=3$, there are $9! = 362880$ permutation operators and the GME-AME measure and the Scott measure is calculated for these states. The GME-AME measure is plotted in Fig~(\ref{E_qutrit}), and the Scott measure is in Fig.~(\ref{scott_qutrit}). It can be seen that there are totally $72$ states are non-GME and also there are $72$ states that are AME states. The GME-AME measure is more robust in clasifying the various classes as there are totally $33$  classes of permutation states according to GME-AME where as totally $16 $ classes are detected through the Scott measure. The exact values and the number of permutation states in these classes are listed in the table (\ref{tab:scott_classes}) for GME-AME measure and in table (\ref{E_qutrit_table}) for Scott measure. 

\begin{table}[t]
    \centering
    \begin{tabular}{|c|c|c| |c|c|c| } 
        \hline
        \textbf{No.} &$\mathbb{S}ct_{2}((\ket{\psi})$ & \textbf{ States} & No & $\mathbb{S}ct_{2}((\ket{\psi})$ & \textbf{States}  \\
        \hline
        1  & 0.6667 & 72    & 9  & 0.8889 & 27432  \\
        2  & 0.8148 & 1170  & 10  & 0.8981 & 36288  \\
        3  & 0.8333 & 864   & 11 & 0.9167 & 101376 \\
        4  & 0.8148 & 1422  & 12 & 0.9352 & 46656  \\
        5  & 0.8796 & 20736 & 13 & 0.8519 & 1296  \\
        6  & 0.8704 & 10368 & 14 & 0.9074 & 44064  \\
        7  & 0.8889 & 27432 & 15 & 0.9259 & 44712 \\
        8 & 0.9630 & 3888  & 16 & 1.0000 & 72\\
        \hline
    \end{tabular}
   
     \caption{Table of $\mathbb{S}ct_2 (\ket{P})$ values of permutation qutrit states ,There are 16 classes of $\mathbb{E}$ permutation states are there out of $9!=362880$ states }
    \label{tab:scott_classes}
\end{table}

\section{Conclusion and discussions \label{sec:conclu}}
In this study, we introduced the GME-AME measure to detect both genuine multipartite entanglement and absolutely maximally entangled states. The measure has a defined range with a lower bound of 0 for non-GME states and an upper bound of 1 for AME states. The GME-AME measure is applied to study four partite states using operator to state mapping for various classes of four party qubit states and permutation states for the qubits and qutrits. We have compared our results with other independent measures of GME and AME and found that the GME-AME measure proposed in this work is robust in classifying the multipartite entangled states. 

The main interesting extensions of the work relate to applying it for a higher number of parties and applying the GME-AME measure in the quantum information theoretic tasks. In the case of many-body systems, it is interesting to see the entanglement growth by considering the entire many-body system as a four-partite system and see the nature of entanglement growth. It is also interesting to see the efficiency of calculating this in the quantum computer for a higher number of parties. The concentratable entanglement is used to classify the multipartite entangled states in a hierarchy~~\cite{schatzki2024hierarchy}. The non-zero value of concentratable entanglement does not guarantee GME; hence, constructing the hierarchy in terms of GME-AME states would give further insights into the multipartite entanglement structure.

\section{acknowledgements}
 SA acknowledges the start-up research grant from
SERB, Department of Science and Technology, Govt. of India (SRG/2022/000467). Authors thank Ranjan Modak, Nripendra Majumdar and Songbo Xie for helpful discussions.

\bibliography{reference,ent_local}

\newpage
\appendix

\section{\label{9 different classes}
9 different classes in 4 party qubit states}
In the four-partite case, there are nine distinct classes of entanglement, where any qubit state of a four-partite system can be transformed into one of these classes via LOCC transformations ~\cite{verstraete2002four}. A detailed investigation of these states reveals that the $\ket{G_{abcd}}$ class corresponds to the same class as the $\ket{X}$ state. Moreover, only the $\ket{L_{0_{3 \bigoplus 1} 0_{3 \bigoplus 1} }}$ state has an GME-AME measure of zero. For the remaining classes, the GME-AME measure is non-zero, indicating that they are genuinely multipartite entangled (GME) states.The following 9 clsses are as :

\begin{widetext}
\begin{align*}
\ket{G_{abcd}} &= \frac{a+d}{2} (|0000\rangle + |1111\rangle) + \frac{a-d}{2} (|0011\rangle + |1100\rangle)
           + \frac{b+c}{2} (|0101\rangle + |1010\rangle) + \frac{b-c}{2} (|0110\rangle + |1001\rangle) \\[10pt]
\ket{L_{abc_2}} &= \frac{a+b}{2} (|0000\rangle + |1111\rangle) + \frac{a-b}{2} (|0011\rangle + |1100\rangle)
           + c (|0101\rangle + |1010\rangle) + |0110\rangle \\[10pt]
\ket{L_{a_2b_2}} &= a (|0000\rangle + |1111\rangle) + b (|0101\rangle + |1010\rangle) + |0110\rangle + |0011\rangle \\[10pt]
\ket{L_{ab_3}} &= a (|0000\rangle + |1111\rangle) + \frac{a+b}{2} (|0101\rangle + |1010\rangle)
           + \frac{a-b}{2} (|0110\rangle + |1001\rangle) + \frac{i}{\sqrt{2}} (|0001\rangle + |0010\rangle)
           + (|0111\rangle + |1011\rangle) \\[10pt]
\ket{L_{a_4}} &= a (|0000\rangle + |0101\rangle + |1010\rangle + |1111\rangle)
          + (i |0001\rangle + |0110\rangle - i |1011\rangle) \\[10pt]
\ket{L_{a_2 \bigoplus 0_{3 \bigoplus 1}}} &= a (|0000\rangle + |1111\rangle) + (|0011\rangle + |0101\rangle + |0110\rangle) \\[10pt]
\ket{L_{0_{5 \bigoplus 3}}} &= (|0000\rangle + |0101\rangle + |1000\rangle) + |1110\rangle \\[10pt]
\ket{L_{0_{7 \bigoplus 1}}} &= (|0000\rangle + |1011\rangle + |1101\rangle + |1110\rangle) \\[10pt]
\ket{L_{0_{3 \bigoplus 1}0_{3 \bigoplus 1}}} &= (|0000\rangle + |0111\rangle) 
\end{align*}
\end{widetext}

\begin{figure}[H]
    \centering
    \includegraphics[width=8cm]{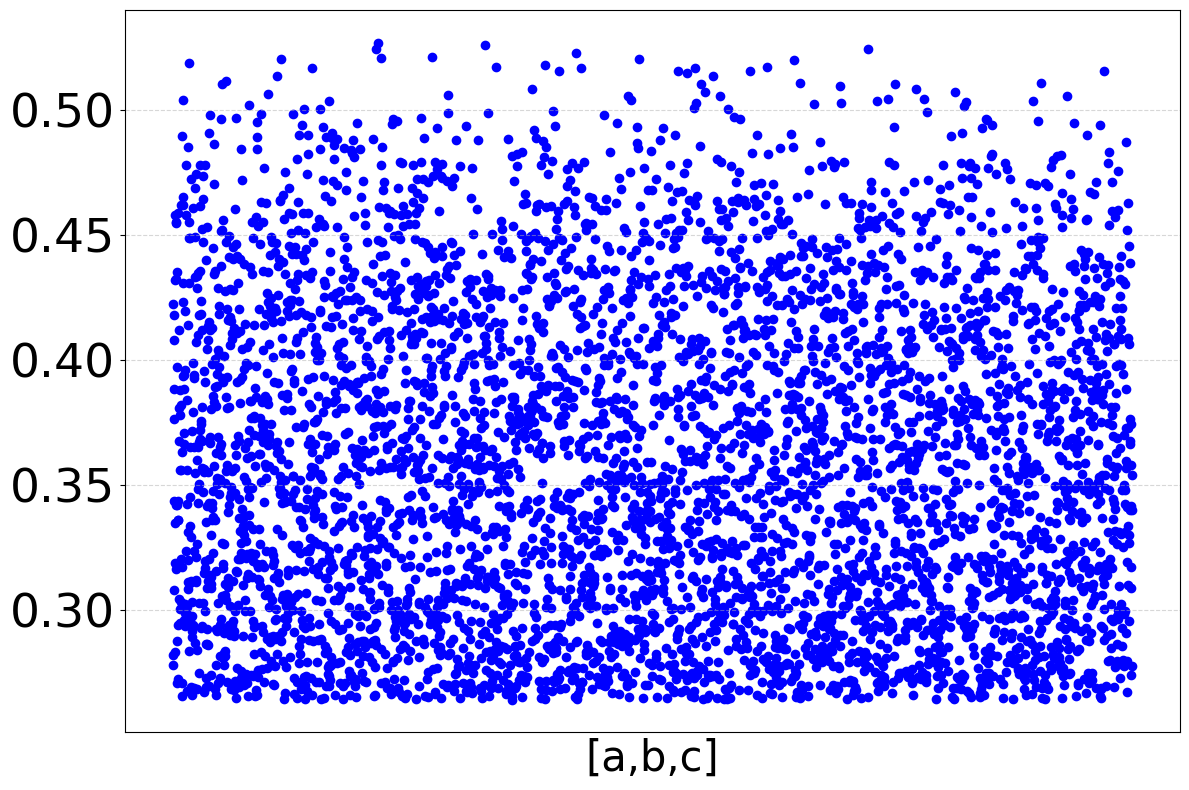}
    \caption{$\mathbb{E}(\ket{L_{abc_{2}}})$ For different 1000 [a,b,c] values}
    \label{second class}
\end{figure}

\begin{figure}[H]
    \centering
    \includegraphics[width=8cm]{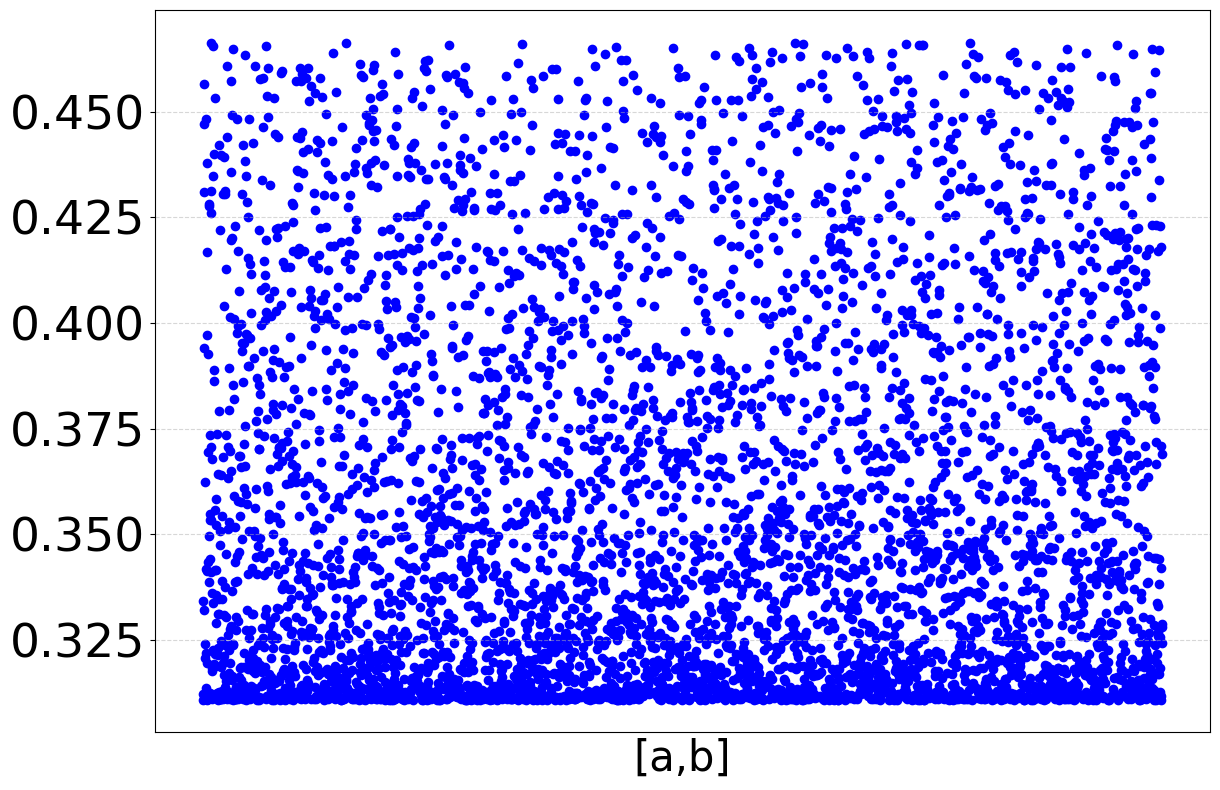}
    
    \caption{$\mathbb{E}(\ket{L_{a_{2}b_{2}}})$ For different 1000 [a,b,c] values}
    \label{third class}
\end{figure}

\begin{figure}[H]
    \centering
    \includegraphics[width=8cm]{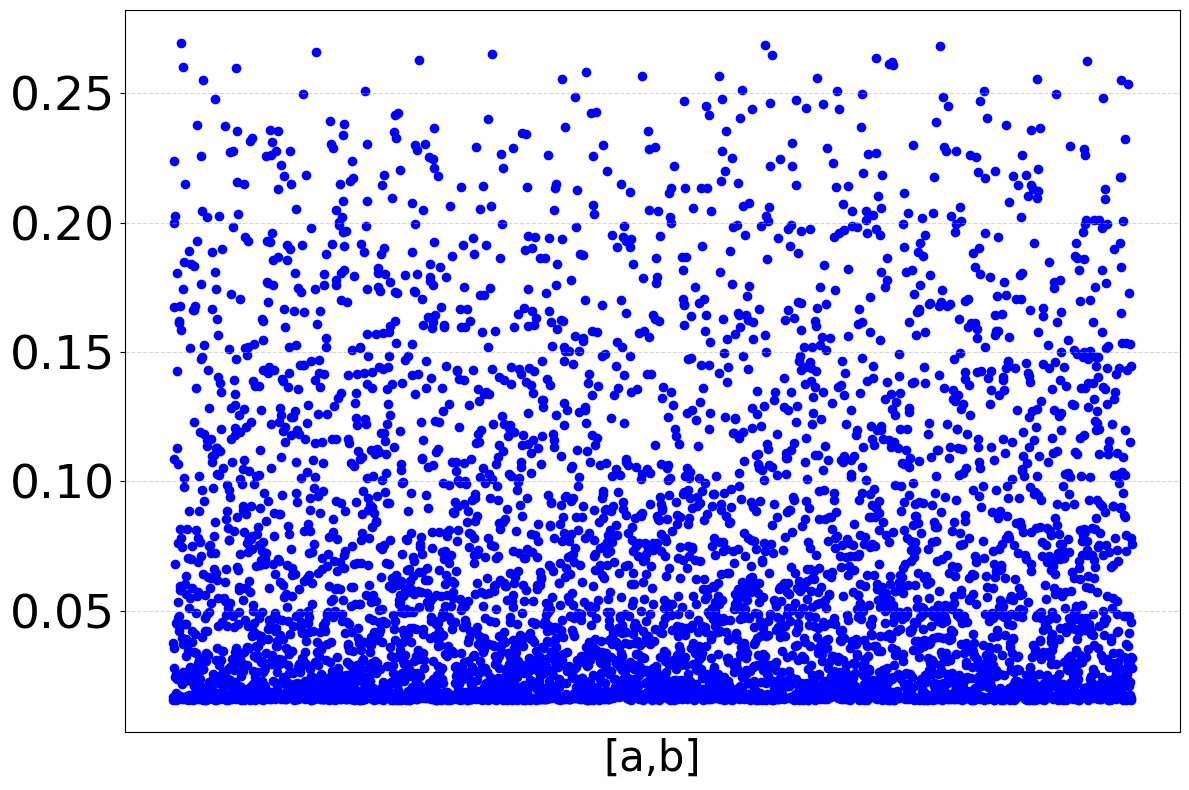}
    
     \caption{$\mathbb{E}(\ket{L_{ab_{3}}})$ For different 1000 [a,b,c] values}
    \label{fourth class}
\end{figure}

 The state $\ket{L_{0_{3 \bigoplus 1}0_{3 \bigoplus 1}}}$ state has an GME-AME measure  zero. For the remaining classes, the GME-AME
 measure is non-zero, indicating that they are GME  states. The GME-AME  $\mathbb{E}$ values for all classes are calculated and given below.For$\ket{L_{abc_2}}$ the GME-AME measure  $\mathbb{E}$ value between 0.25 and 0.55 Fig.~\ref{second class},$\ket{L_{a_2b_2}}$ ranges from 0.3 to 0.475 Fig ~\ref{third class} and $\ket{L_{a_2b_2}}$ from 0.03 to 0.26,Fig ~\ref{fourth class}. The calculated values of $\mathbb{E}(\ket{L_{a_4}}) = 0.4,\mathbb{E}(\ket{L_{a_2 0_{3 \bigoplus 1}}})= 0.23 ,\mathbb{E}(\ket{L_{0_{5 \bigoplus 3}}}) = 0.156,  \mathbb{E} (\ket{L_{0_{7 \bigoplus 1}}}) = 0.435,\mathbb{E} (\ket{L_{0_{3 \bigoplus 1} 0_{3 \otimes 1}}}) 
 = 0$

\section{Enphasing of Permutation states}
In this Section, we provide the plots for the Scott measure and Polygon measure for the enphasing of qubit permutation states. 
\begin{figure}[H]
    \centering
    \includegraphics[width=8cm]{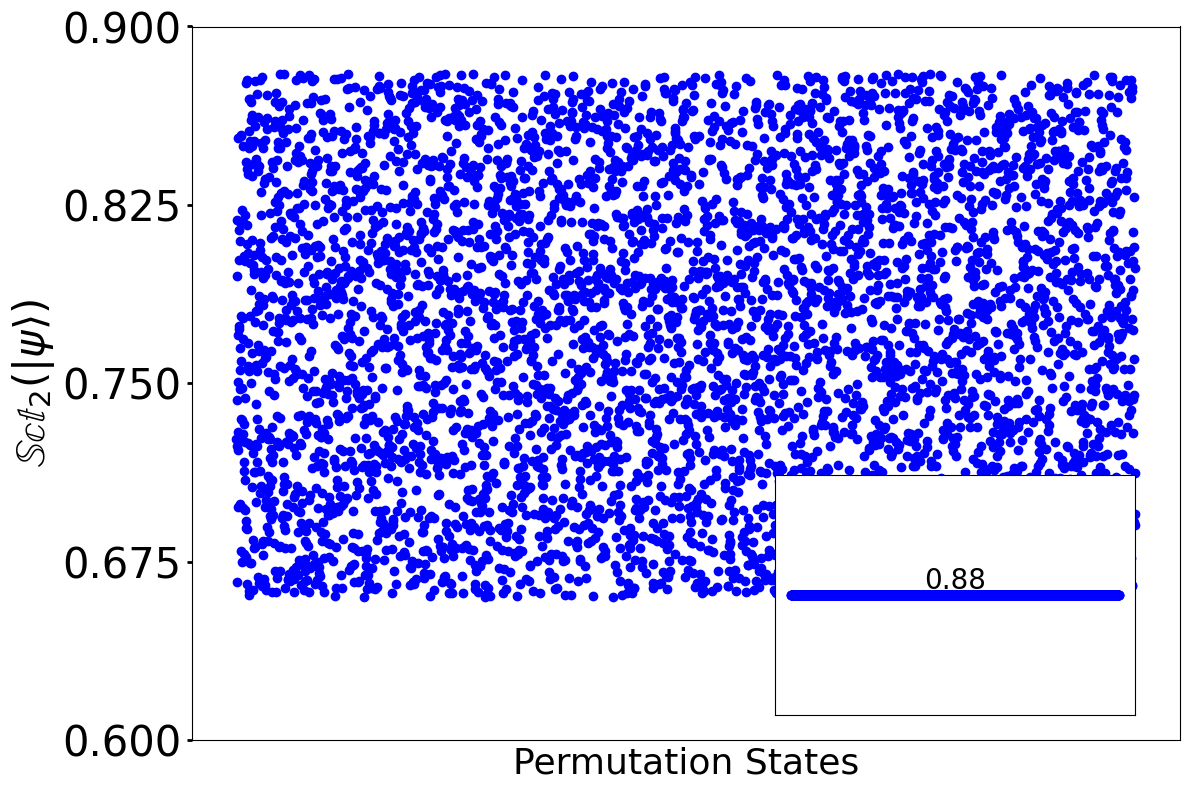}
    \caption{$\mathbb{S}ct_{2}\ket{P}$ enphasing qubit permutation state.}
    \label{scott enphasing}
\end{figure}

\begin{figure}[H]
    \centering
    \includegraphics[width=8cm]{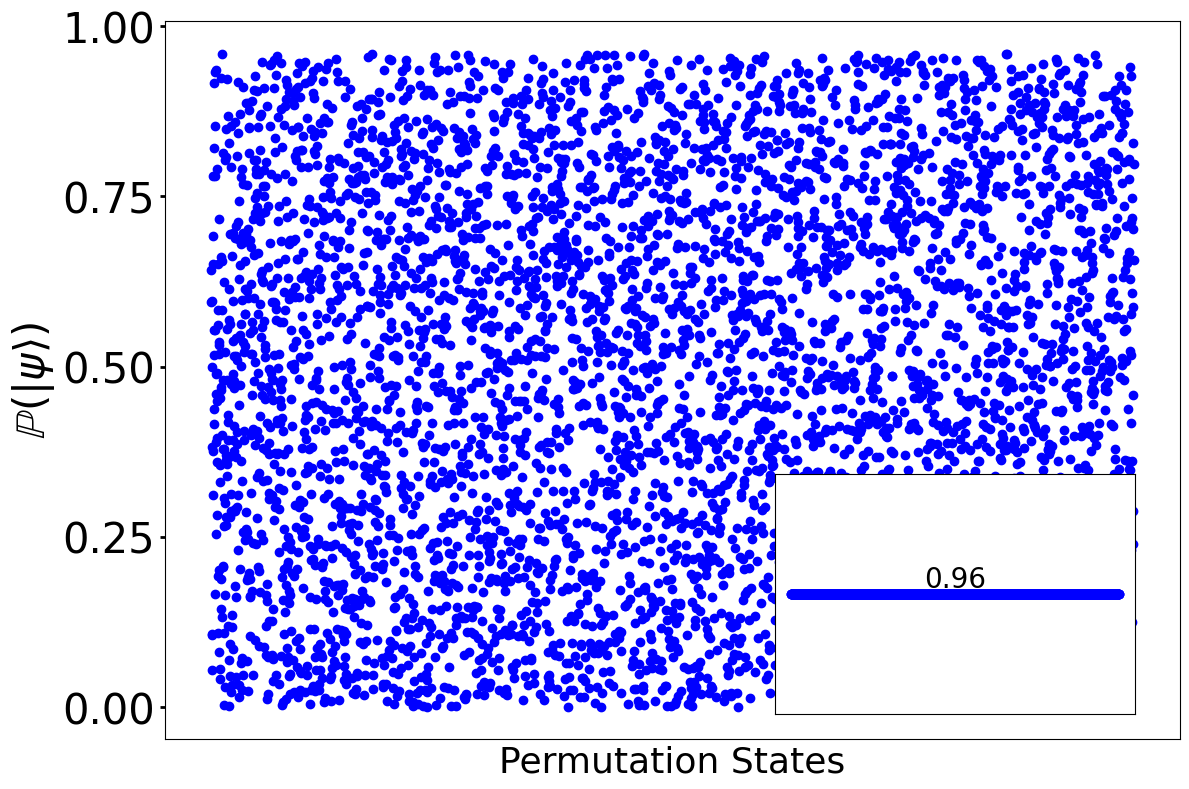}
    \caption{The polygon measure of $\mathbb{P}(\ket{P})$ enphased qubit permutation states.}
    \label{polygon enphasing}
\end{figure}
\end{document}